\documentclass[10pt,journal,draftcls,onecolumn]{IEEEtran}

\usepackage{graphicx}

\ifCLASSOPTIONcompsoc
\else
\fi

\graphicspath{{FIG/}}
\begin{document}
%
\title{Constructing LDPC Codes from Partition and Latin-Style Splicing}

%
%
%
%

\author{Guohua~Zhang,~\IEEEmembership{}
        Yulin~Hu,~\IEEEmembership{}
        and~Qinwei~He~\IEEEmembership{}
\IEEEcompsocitemizethanks{\IEEEcompsocthanksitem Guohua~Zhang is with the China Academy of Space Technology, Xi'an, People's Republic of China. 
Email: zhangghcast@163.com
\IEEEcompsocthanksitem Yulin~Hu and Qinwei~He are with the Institute for Theoretical Information Technology, RWTH Aachen university, Aachen, Germany.}%
}

\IEEEtitleabstractindextext{%
\begin{abstract}
A novel method guaranteeing nondecreasing girth is presented for constructing longer low-density parity-check (LDPC) codes from shorter ones. The parity-check matrix of a shorter base code is decomposed into $N~(N\geq 2)$ non-overlapping components with the same size. Then, these components are combined together to form the parity-check matrix of a longer code, according to a given $N\times N$ Latin square. To illustrate this method, longer quasi-cyclic (QC) LDPC codes are obtained with girth at least eight and satisfactory performance, via shorter QC-LDPC codes with girth eight but poor performance. The proposed method naturally includes several well-known methods as special cases, but is much more general compared with these existing approaches.
\end{abstract}

\begin{IEEEkeywords}
LDPC codes, Quasi-cyclic, Girth, Latin square, Greatest common divisor
\end{IEEEkeywords}}

\maketitle

\IEEEdisplaynontitleabstractindextext
\section{Introduction} QC-LDPC codes have gained significant attention, due to their strength of facilitating simple encoders and decoders. It is generally accepted \cite{F04}\cite{LK04} that girth is one important factor (among others), affecting performance of LDPC codes.
There have been a large number of \emph{specific} methods for constructing QC-LDPC codes with decent or large girth. However, except for a couple of methods, such as \cite{KLF01}\cite{MMY07}\cite{MYP06}, \emph{general} skills to obtain longer codes with nondecreasing girth from shorter ones are rarely available. In this Letter, a new such method is proposed to design longer compound QC-LDPC codes from shorter base codes, while the girth of former is equal to or greater than that of the latter. On the basis of some shorter codes (from greatest-common-divisor, viz.,  GCD method \cite{ZSW13}) with girth eight but unsatisfactory performance, the resultant longer codes not only possess a girth at least eight but also outperform some well-known specific classes of QC-LDPC codes. The strength of the new method lies in simplicity (using only simple operations of partition and splicing) and flexibility (applicable to any base code).

\section{Preliminaries} An LDPC code is defined as the null space of a sparse parity-check matrix (PCM). If a PCM has a constant column (resp. row) weight of $J$ (resp. $L$), then it yields a $(J,L)$-regular code. Generally, the PCM of a QC-LDPC code is an array of circulants, which may include zero matrix (ZM), circulant permutation matrix (CPM) or summation of distinct CPMs.
An LDPC code with PCM $\textbf{H}$ can be represented by its associated Tanner graph, $TG(\textbf{H})$, and the length of the shortest cycle in the graph is called girth (or girth of the code/PCM). Denote by $g(\textbf{H})$ the girth of $\textbf{H}$. If $\textbf{H}$ is an array of $P\times P$ CPMs/ZMs, then it can be completely determined by $P$ and an exponent matrix $\textbf{E}$ with entries in the set $\{\infty, 0,1,\cdots,P-1\}$, where $\infty$ corresponds to a $P\times P$ ZM and any other entry (say $e$) to a $P\times P$ identity matrix with rows cyclically shifted to the right by $e~(mod~P)$ positions \cite{WZZYS}. For this case, $\textbf{H}$ and its girth can also be denoted by $\textbf{H}(\textbf{E},P)$ and $g(\textbf{E},P)$, respectively.

Let $\textbf{Z}_N=\{0,1,\cdots,N-1\}$. A Latin square of order $N$ \cite{Zhang16} is an $N\times N$ array in which each cell contains a symbol from $\textbf{Z}_N$, such that each
symbol occurs exactly once in each column and each row.

\section{A General Method from Partition and Latin-Style Splicing (PS)} The base PCM $\textbf{H}_0$ is assumed to be an $m\times n$ array of $P\times P$ matrices over $GF(2)$. Clearly, the PCM of a QC-LDPC code is a special case of $\textbf{H}_0$.
Given an integer $N\geq 2$, select $N$ masking matrices $\textbf{M}_k~(0\leq k\leq N-1)$ of size $m\times n$ defined over the integer set $\{0,1\}$, such that the ordinary summation $\sum_{k=0}^{N-1} \textbf{M}_k$ equals an $m\times n$ all-one matrix. Let $\textbf{A}=[a_{i,j}]~(0\leq i,j\leq N-1)$ be a Latin square of order $N$.
From $\textbf{H}_0$, a new $mPN\times nPN$ PCM \textbf{H} can be obtained by
\begin{equation}
\left[
  \begin{array}{cccc}
    \textbf{H}_0\otimes f(\textbf{M}_{a_{0,0}})&  \textbf{H}_0\otimes f(\textbf{M}_{a_{0,1}}) & \cdots &  \textbf{H}_0\otimes f(\textbf{M}_{a_{0,N-1}})\\
    \textbf{H}_0\otimes f(\textbf{M}_{a_{1,0}})&  \textbf{H}_0\otimes f(\textbf{M}_{a_{1,1}}) & \cdots &  \textbf{H}_0\otimes f(\textbf{M}_{a_{1,N-1}})\\
    \vdots & \vdots & \ddots & \vdots\\
    \textbf{H}_0\otimes f(\textbf{M}_{a_{N-1,0}})&  \textbf{H}_0\otimes f(\textbf{M}_{a_{N-1,1}}) & \cdots &  \textbf{H}_0\otimes f(\textbf{M}_{a_{N-1,N-1}})\\
   \end{array}
\right]
\end{equation}
where $f(\textbf{M})$ maps each 0 (resp. 1) in $\textbf{M}$ to a $P\times P$ zero (resp. all-one) matrix over $GF(2)$, and $\otimes $ is an element-by-element multiplication
defined by $x \otimes y=1$ if and only if $x=y=1$.

\emph{Theorem 1}: $g(\textbf{H})\geq g(\textbf{H}_0)$.

\emph{Proof}: According to the definitions of \textbf{M}'s and Latin square \textbf{A}, two adjacent edges within $TG(\textbf{H})$ correspond to two adjacent edges within $TG(\textbf{H}_0)$. Therefore, if there is a cycle of length $2l$ in $TG(\textbf{H})$, then there must exist a corresponding cycle with the same length in $TG(\textbf{H}_0)$. This completes the proof.

It is easily seen that the row/column weight distribution of $\textbf{H}$ is the same as that of $\textbf{H}_0$, and the designed rate for $\textbf{H}$ equals that for $\textbf{H}_0$. For $P=1$, $\textbf{H}_0$ corresponds to an LDPC code not necessarily with a special structure, that is to say, $\textbf{H}_0$ can be an arbitrary binary matrix. This suggests that the new method is applicable to any base code. If $\textbf{H}_0$ is an $m\times n$ array of $P\times P$ CPMs/ZMs, then \textbf{H} (Equ.1) can be described by (via its exponent matrix)
\begin{equation}
\textbf{E}=
\left[
  \begin{array}{llll}
    \textbf{E}_0\tilde \otimes \textbf{M}_{a_{0,0}}&  \textbf{E}_0\tilde \otimes \textbf{M}_{a_{0,1}} & \cdots &  \textbf{E}_0\tilde \otimes \textbf{M}_{a_{0,N-1}}\\
    \textbf{E}_0\tilde \otimes \textbf{M}_{a_{1,0}}&  \textbf{E}_0\tilde \otimes \textbf{M}_{a_{1,1}} & \cdots &  \textbf{E}_0\tilde \otimes \textbf{M}_{a_{1,N-1}}\\
    \vdots & \vdots & \ddots & \vdots\\
    \textbf{E}_0\tilde \otimes \textbf{M}_{a_{N-1,0}}&  \textbf{E}_0\tilde \otimes \textbf{M}_{a_{N-1,1}} & \cdots &  \textbf{E}_0\tilde \otimes \textbf{M}_{a_{N-1,N-1}}\\
   \end{array}
\right]
\end{equation}
where $\textbf{E}_0$ is the exponent matrix of $\textbf{H}_0$.
The notation $\tilde \otimes$ stands for an element-by-element operation defined by $x \tilde \otimes~0=\infty$, and $x \tilde \otimes~1=x$, where $x\in \{\infty, 0,1,\cdots,P-1\}$.

\emph{A Special Case}: Let $N=2$ and $n=Km$. Suppose that both $\textbf{M}_0$ and $\textbf{M}_1$ are $1\times K$ arrays of $K$ identical $m\times m$ matrices. Let $\textbf{M}_0=[\textbf{X},\cdots,\textbf{X}]$, where each element in the lower-triangle (including diagonal) of \textbf{X} is '1', and '0' elsewhere. Define $\textbf{M}_1=\textbf{1}_{m\times n}-\textbf{M}_0$. Then, $\textbf{E}$ can be obtained from $\textbf{E}_0$ by
\begin{equation}
\left[
  \begin{array}{llll}
    \textbf{E}_0\tilde \otimes \textbf{M}_{a_{0,0}} &  \textbf{E}_0\tilde \otimes \textbf{M}_{a_{0,1}}\\
     \textbf{E}_0\tilde \otimes \textbf{M}_{a_{1,0}} & \textbf{E}_0\tilde \otimes \textbf{M}_{a_{1,1}}\\
   \end{array}
\right],
\textbf{A}=
\left[
  \begin{array}{ll}
  a_{0,0} & a_{0,1}\\
  a_{1,0} & a_{1,1}\\
   \end{array}
\right]=
\left[
  \begin{array}{ll}
  0 & 1\\
  1 & 0\\
   \end{array}
\right]
\end{equation}

Clearly, the above \textbf{E} is equivalent to that investigated in \cite{TLZL06} (Section VII). It is pointed out \cite{TLZL06} that $g(\textbf{E},P)$ is ensured to be at least 6, provided $g(\textbf{E}_0,P)\geq 6$. The method in this Letter, however, is more general in the sense that $N$ is not limited to 2, and $n$ and $m$ can be selected arbitrarily. Moreover, given an arbitrary $\textbf{E}_0$ with $g(\textbf{E}_0,P)=2g_0$, Theorem 1 guarantees an exponent matrix \textbf{E} with $g(\textbf{E},P)\geq 2g_0$.

\emph{Example 1}: Given an exponent matrix
 \begin{equation}
\textbf{E}_0=
\left[
  \begin{array}{llll}
0 & 0 & 0& 0\\
0 & 1 & 3& 4\\
0 & 2 & 6& 5\\
\end{array}
\right]
\end{equation}
it is easily checked that $g(\textbf{E}_0,P)=4$ for any $P\geq 7$, as there is an equation $(1-2)+(5-4)=0~(mod~P)$ regardless of $P$. Set $N=2$, and define
\begin{equation}
\textbf{M}_0=
\left[
  \begin{array}{llll}
1 & 1 & 1& 1\\
1 & 1 & 1& 1\\
1 & 0 & 0& 1\\
\end{array}
\right],
\textbf{A}=
\left[
  \begin{array}{ll}
a_{0,0} & a_{0,1}\\
a_{1,0} & a_{1,1}\\
\end{array}
\right]
=
\left[
  \begin{array}{ll}
0 & 1\\
1 & 0\\
\end{array}
\right]
\end{equation}
Then, $\textbf{M}_1=\textbf{1}_{3\times 4}-\textbf{M}_0$.
Thus, the new method yields
\begin{equation}
\textbf{E}=
\left[
  \begin{array}{ll}
      \textbf{E}_0\tilde \otimes \textbf{M}_{0} & \textbf{E}_0\tilde \otimes \textbf{M}_{1}\\
      \textbf{E}_0\tilde \otimes \textbf{M}_{1} & \textbf{E}_0\tilde \otimes \textbf{M}_{0}\\
\end{array}
\right]
=
\left[
  \begin{array}{llll|llll}
     0 &0 &  0 &  0 &   &  &  & \\
     0 &1 &  3&  4 &  &  &   & \\
     0 &  &  &  5 &  &   2 &  6  & \\
  \hline
     &  &   &  & 0 & 0 & 0 &   0\\
    &   &   & &  0 &  1 &  3 &   4\\
     &   2 &  6 & & 0&  &  &    5\\
\end{array}
\right]
\end{equation}
It is readily verified that $g(\textbf{E},P)=8$ for any $P\geq 7$.
\section{Comparison with existing methods} There exist some general methods to construct longer compound LDPC codes with nondecreasing girth from one or several shorter base code(s). The \emph{column splitting} method \cite{KLF01} is capable of improving on the flexibility of code length and increase code rate by splitting each column of a PCM with large column weight into several columns. The method in \cite{MMY07} yields a longer QC-LDPC codes with girth at least $2g$~($2g=6$ or $8$) from two shorter QC-LDPC codes both with girth at least $2g$. This method extends the row weight of QC-LDPC code (hence code rate increased), but only ensure a girth not exceeding eight even if both base codes have a girth larger than eight. The Chinese-remainder-theorem (CRT) method \cite{MYP06} can be employed to construct longer QC-LDPC codes by combining two shorter base QC-LDPC codes. The girth of the resultant code is not smaller than the maximal girth of the two base codes. Obviously, the new method is different from all the aforementioned methods.

Besides, if the Latin square $\textbf{A}$ is selected as $[a_{i,j}]=[i-j~(mod~N)]$, then Equ. (1) is reduced to the PCM of spatially coupled LDPC block codes \cite{MLC15}. The base matrices of the terminated ensembles and the tail-biting counterpart (Equ. (8) and Equ. (10) in \cite{MLC15}) are the same as the partial matrix and the whole matrix of \textbf{E}, respectively. However, since \textbf{A} generally possesses many forms different from the above choice, \textbf{E} and \textbf{H} are generally not identical to those of the spatially coupled LDPC codes.

\section{Simulations} The application of the new general method is illustrated by several examples. Set $\textbf{A}$ as $[a_{i,j}]=[i-j~(mod~N)]$. Given a matrix $\textbf{M}_0$, set $\textbf{M}_1=\textbf{1}_{m\times n}-\textbf{M}_0$, and let $\textbf{M}_k=\textbf{O}~(2\leq k\leq N-1)$ if $N\geq 3$. The matrix $\textbf{M}_0$ can be selected by different ways, in which three options are considered in this Letter. (a) For diagonal (D) partition, let $\textbf{M}_0=[\textbf{X},\cdots, \textbf{X}]$, where $\textbf{X}$ is an $m\times m$ matrix with 0's in diagonal and 1's elsewhere. (b) For triangle (T) partition, $\textbf{M}_0=[\textbf{X},\cdots, \textbf{X}]$, where $\textbf{X}(i,j)=1$ for $0\leq j\leq i~(0\leq i\leq m-1)$ and 0 elsewhere; and (c) For Hamming-like (H) partition, $\textbf{M}_0$ is a 0-1 $m\times n$ matrix in which the columns are as distinct as possible. For simulations, a BI-AWGN channel with BPSK modulation and SPA with 50 iterations are assumed.

\emph{Example 2}: From the GCD construction \cite{ZSW13}, $\textbf{E}_0=[0,1,L,L+1]^T\cdot[0,1,\cdots, L-1]~(mod~P)$ is chosen as the base exponent matrix, where $P=64$ and $L=8$. Set $N=4$. For H partition, $\textbf{M}_0$ can be selected as
\begin{equation}
\textbf{M}_0=
\left[
  \begin{array}{llllllll}
  1 & 0& 1& 1& 1 &0 & 0 & 0\\
  1 & 1& 0& 1& 0 &1 & 0 & 0\\
  1 & 1& 1& 0& 0 &0 & 1 & 0\\
  0 & 1& 1& 1& 0 &0 & 0 & 1\\
\end{array}
\right]
\end{equation}
From D,T and H partitions, three PS $(4,8)$-regular QC-LDPC codes are obtained with girth at least eight. For comparison purpose, a girth-8 GCD code and a girth-6 quad.-cong. code, both (4,8)-regular, are generated. The GCD code is obtained by setting $\textbf{E}=[0,1,L,L+1]^T\cdot[0,1,\cdots,L-1]~(mod~P)$ where $P=256$ and $L=8$, and the quad.-cong. code is randomly constructed by the method \cite{HHY08} with a prime CPM size $257$. From Fig.1, it is observed that the GCD code performs the worst, the PS-D and PS-T codes better, and the PS-H code the best, which can be partly explained by the fact that during the simulation, codewords with weight 14, 16, 16 are found for codes from GCD, PS-D and PS-T methods, respectively, and no codewords with small weight occur for the PS-H code. Moreover, the PS-H code noticeably outperform the well-known random quad.-cong. code.

\begin{figure}[htb]
  \centering
  \includegraphics[width=10cm,bb=85 270 480 570]{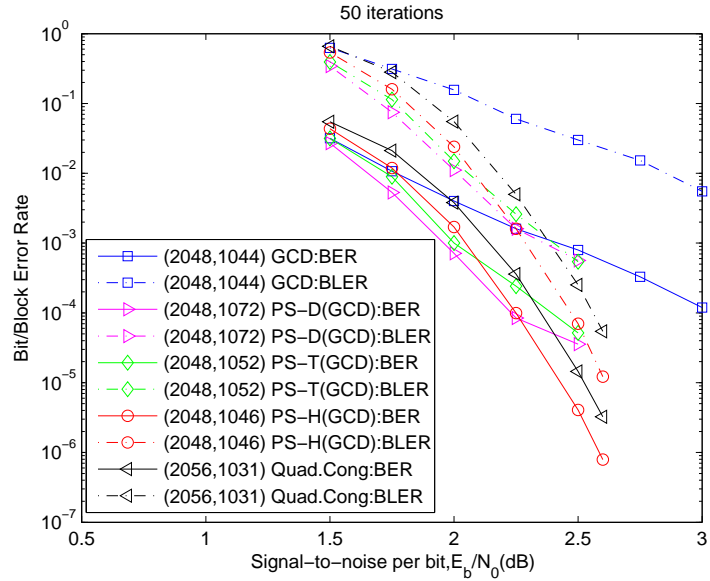}
  \caption{Performance comparison of (4,8)-regular QC-LDPC codes from PS, GCD and Quad Cong. methods}
  \label{f4}
\end{figure}

\begin{figure}[htb]
  \centering
  \includegraphics[width=10cm,bb=85 270 480 570]{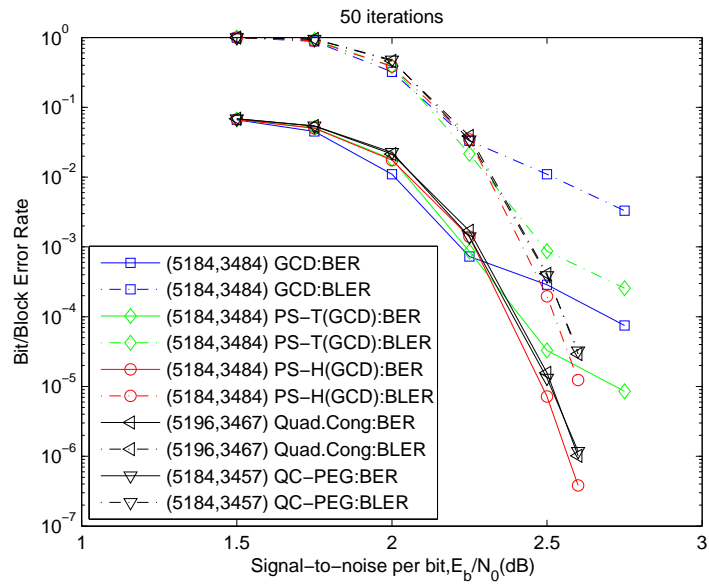}
  \caption{Performance comparison of (4,12)-regular QC-LDPC codes from PS, Quad Cong. and QC-PEG methods}
  \label{f7x}
\end{figure}

\emph{Example 3}: According to the GCD construction \cite{ZSW13}, set $\textbf{E}_0=[0,1,L,L+1]^T\cdot[0,1,\cdots, L-1]~(mod~P)$ as the base exponent matrix, where $P=144$ and $L=12$. Let $N=3$, and for H partition set
\begin{equation}
\textbf{M}_0=
\left[
  \begin{array}{llllllllllll}
     1  &   0   &  1   &  1   &  1  &   0  &   0  &   1 &    1 &    0 &  0 &  0\\
     1  &   1   &  0   &  1   &  1  &   1  &   0  &   0 &    0 &    1 &  0 &  0\\
     1  &   1   &  1   &  0   &  0  &   1  &   1  &   0 &    0 &    0 &  1 &  0\\
     0  &   1   &  1   &  1   &  0  &   0  &   1  &   1 &    0 &    0 &  0 &  1\\
\end{array}
\right]
\end{equation}
From T and H partitions, two PS $(4,12)$-regular QC-LDPC codes are obtained with girth at least eight. A (4,12)-regular girth-8 GCD code is also generated by the exponent matrix $[0,1,L,L+1]^T\cdot[0,1,\cdots, L-1]~(mod~P)$, where $P=432$ and $L=12$. A (4,12)-regular quad.-cong code is randomly generated with girth 6; besides, a (4,12)-regular QC-PEG code \cite{LK04} is obtained with girth 8, the PCM of which is a $12\times 36$ array composed of 142 CPMs, 289 ZMs and one summation of two CPMs. We observe in Fig.2 that the PS-H code performs better than the quad.-cong code and the QC-PEG code, while the GCD and PS-T counterparts both suffer from error floor partly due to their relatively poor distance property (during the simulation, codewords with weight 14 and 12 exist for the GCD code and PS-T code, respectively, and no codewords with small weight is found for the PS-H code).

\section{Conclusion} By non-overlapping partition and Latin-square-style splicing, a general method (PS) preserving girth is proposed to yield longer LDPC codes from shorter base ones. Numerical results show that the codes generated by combining PS method (using Hamming-like partition) with some poor base codes perform very well compared with the well-known QC-PEG and quad.-cong. codes. Finally, it should be pointed out that by applying the new method to some explicitly constructed base codes with large girth (e.g. \cite{KNCS06}, \cite{GM12}), type-1 QC-LDPC codes with large girth such as \cite{WZZYS}\cite{ZM17} can be easily constructed.

%
\IEEEpeerreviewmaketitle


%

\ifCLASSOPTIONcompsoc
  \section*{Acknowledgments}
\else
  \section*{Acknowledgment}
\fi

This work was supported by the National Natural Science Foundation of China under grant 61471294.

\ifCLASSOPTIONcaptionsoff
  \newpage
\fi

\end{document}